\documentclass[conference]{IEEEtran}
\pdfoutput=1
\usepackage{cite}
\usepackage[pdftex]{graphicx}
\graphicspath{{figures/}}
\DeclareGraphicsExtensions{.pdf,.jpeg,.png}
\usepackage{amsmath}

\usepackage{url}

\def\TinBiNN{TinBiNN}
\def\bcerrorfloat32{9.9}
\def\onecaterrorfloat32{0.4}
\def\onecaterrorint8{0.4}
\def\tencaterrorZCAfloat32{11.8}
\def\tencaterrorNOZCAfloat32{13.6}
\def\tencaterrorint8{13.6}
\def\tencatruntime{1,315ms}

\def\onecatruntime{195ms}
\def\inteltencatruntime{6.4ms}
\def\intelonecatruntime{2.0ms}

\begin{document}

\title{TinBiNN: Tiny Binarized Neural Network\\ Overlay in about 5,000 4-LUTs and 5mW}

\author{\IEEEauthorblockN{
Guy G.F. Lemieux\IEEEauthorrefmark{1},
Joe Edwards\IEEEauthorrefmark{1},\\
Joel Vandergriendt\IEEEauthorrefmark{1}, 
Aaron Severance\IEEEauthorrefmark{1},
Ryan De~Iaco\IEEEauthorrefmark{1}\\
Abdullah Raouf\IEEEauthorrefmark{2},
Hussein Osman\IEEEauthorrefmark{2},
Tom Watzka\IEEEauthorrefmark{2},
Satwant Singh\IEEEauthorrefmark{2}}
\IEEEauthorblockA{\IEEEauthorrefmark{1}VectorBlox Computing (firstname.lastname@vectorblox.com)}
\IEEEauthorblockA{\IEEEauthorrefmark{2}Lattice Semiconductor (firstname.lastname@latticesemi.com)}
}

\maketitle

\begin{abstract}
Reduced-precision arithmetic improves the size, cost, power and performance of
neural networks in digital logic.  In convolutional neural networks, the use of
1b weights can achieve state-of-the-art error rates while eliminating
multiplication, reducing storage and improving power efficiency.  The
BinaryConnect binary-weighted system, for example, achieves \bcerrorfloat32\%
error using floating-point activations on the CIFAR-10 dataset.  In this paper,
we introduce \TinBiNN, a lightweight vector processor overlay for accelerating
inference computations with 1b weights and 8b activations.  The overlay is very
small -- it uses about 5,000 4-input LUTs and fits into a low cost iCE40
UltraPlus FPGA from Lattice Semiconductor. To show this can be useful, we build
two embedded `person detector' systems by shrinking the original BinaryConnect
network.  The first is a 10-category classifier with a 89\% smaller network
that runs in \tencatruntime\ and achieves \tencaterrorint8\% error.  The other
is a 1-category classifier that is even smaller, runs in \onecatruntime, and
has only \onecaterrorint8\% error.  In both classifiers, the error can be
attributed entirely to training and not reduced precision.

\end{abstract}

%

\IEEEpeerreviewmaketitle

\section{Description}

State-of-the-art machine learning in computer vision uses Convolutional Neural
Networks (CNNs), where the compute kernel is a 2D convolution (often $3\times 3$). This
computation is repeated at every position in the input map, producing an output
map of similar size. The outputs become inputs for the next layer, and the
process is repeated.

CNNs typically use floating-point data types when computing with GPUs.
BinaryConnect~\cite{courb2015nips} is a new training method that uses 1b
weights to represent $\pm 1$, but still computes with floating-point data.
This saves memory storage and bandwidth, and replaces all multiplications with
conditional negation.  BinaryConnect reportedly achieved \bcerrorfloat32\% test
error rate on CIFAR-10~\cite{CIFAR10}; our reproduction achieved 10.3\% error.
The error is measured as the fraction of test set images that are incorrectly
categorized.

\begin{figure}[t]
    \centerline{\includegraphics[width=0.65\columnwidth]{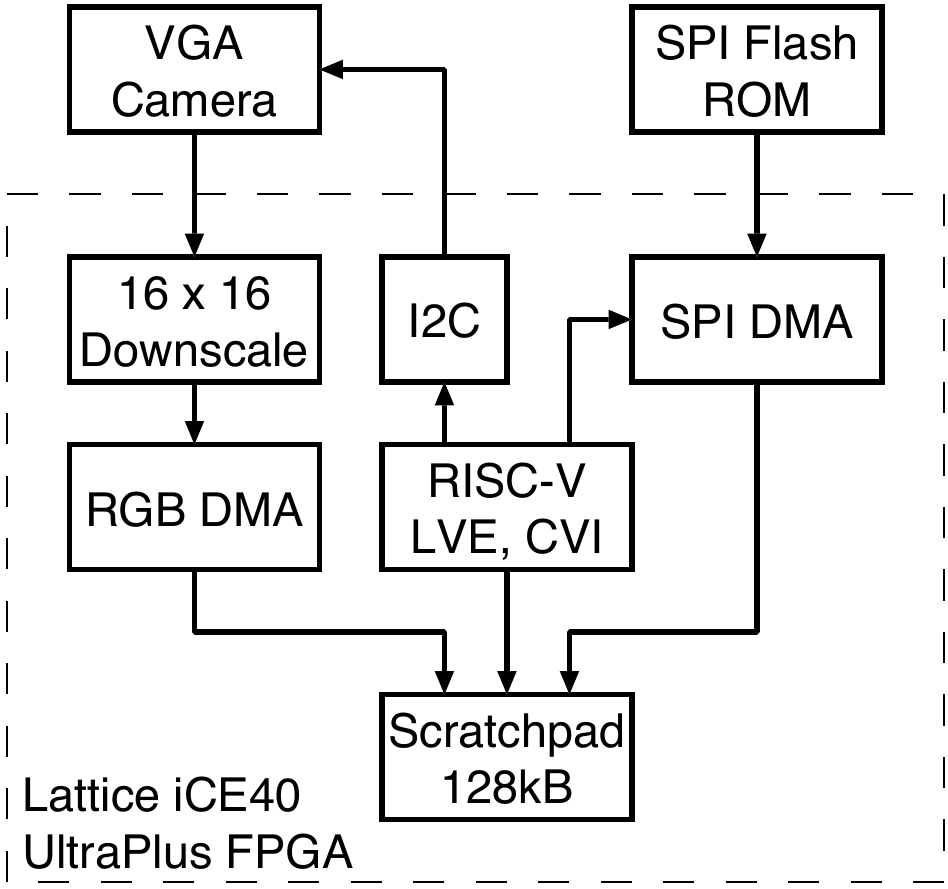}}
    \caption{System diagram\label{fig:system}}
\end{figure}

\begin{figure}[t]
  \begin{centering}
    \includegraphics[width=1.0\columnwidth]{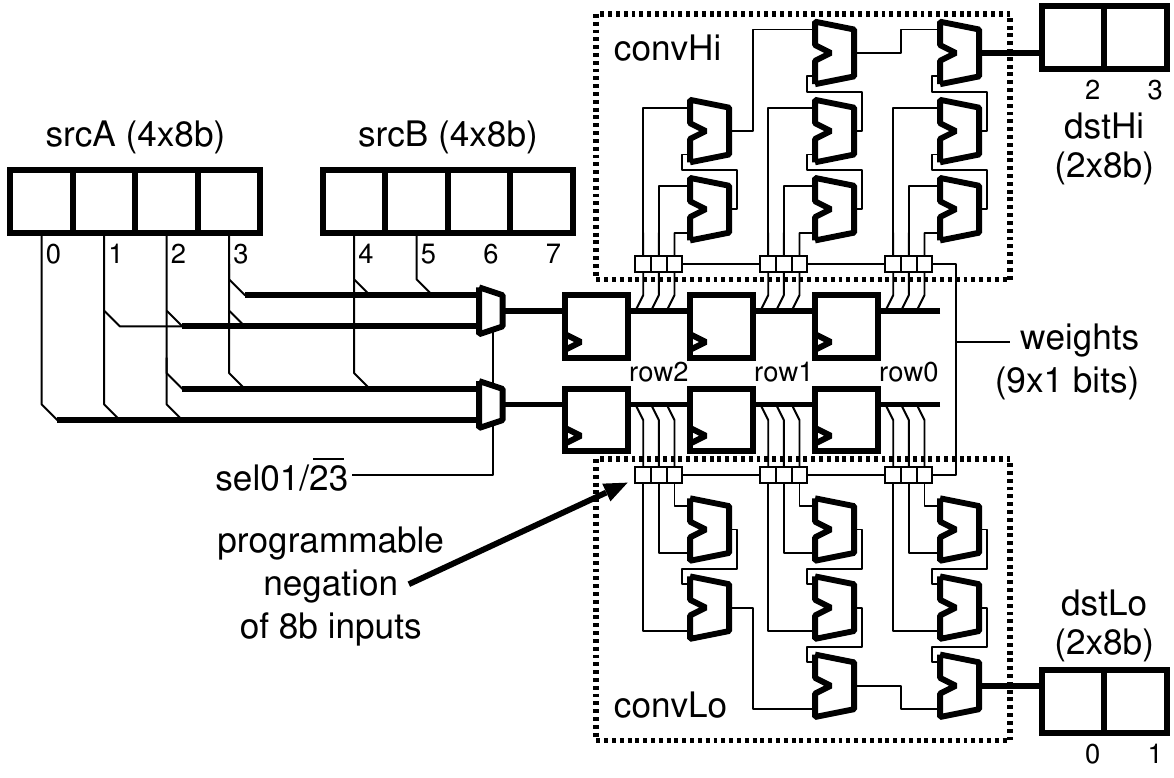}
    \par\end{centering}
    \caption{Binarized CNN custom vector instruction\label{fig:cvi}}
\end{figure}

\begin{figure*}[t]
  \begin{centering}
    \includegraphics[width=2.1\columnwidth]{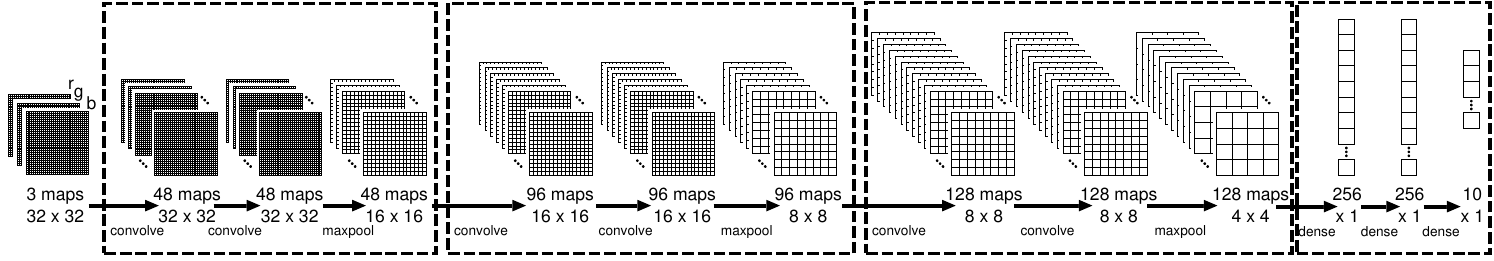}
    \par\end{centering}
    \caption{Reduced binarized CNN structure with 89\% less computation, giving \tencaterrorint8\% error on CIFAR-10\label{fig:network}}
\end{figure*}

In this paper we make three contributions.  First, we optimized the
BinaryConnect system by reducing the network size and computed precision.
We reduced the network size from:\\
\indent
{\tiny 
\ $(2\times 128C3)$-\mbox{MP2}-$(2\times 256C3)$-\mbox{MP2}-$(2\times 512C3)$-\mbox{MP2}-$(2\times 1024\mbox{FC})$-$10$\mbox{SVM}}\\
to:\\
\indent
{\tiny
$(2\times 48C3)$-\mbox{MP2}-$(2\times 96C3)$-\mbox{MP2}-$(2\times 128C3)$-\mbox{MP2}-$(2\times 256\mbox{FC})$-$10$\mbox{SVM}}\\
where $C3$ is a $3\times 3$ ReLU convolution layer, $\mbox{MP2}$ is a $2\times
2$ max-pooling layer, $\mbox{FC}$ is a fully connected layer, and $\mbox{SVM}$
is a L2-SVM output layer.  The new network, shown in Figure~\ref{fig:network},
has 89\% fewer operations than the BinaryConnect reproduction and achieved
\tencaterrorZCAfloat32\% error on CIFAR-10.  For performance, we also dropped
ZCA whitening, increasing error to \tencaterrorNOZCAfloat32\%.  As well, we
converted all computation to fixed point, with hidden layer outputs
(\textit{activations}) using 8b unsigned integers and intermediate sums using
16b and 32b signed integers, maintaining the same error rate of
\tencaterrorint8\%.

Second, for performance, we implemented a hardware accelerator for CNNs with
binary weights and 8b inputs, shown in Figure~\ref{fig:cvi}.  The accelerator
computes two overlapping convolutions in parallel.  In use, input data is
fetched down a column, accepting 8 consecutive bytes each cycle as its two 32b
operands. 
Two passes over the same column are made. The first pass computes two 16b
output convolutions starting at byte offsets 0 and 1 of the input column.  The
second pass computes two more outputs at byte offsets 2 and 3. After that, the
input column advances by 4 bytes and maintains alignment.

Third, for performance and flexibility, we developed the TinBiNN overlay
system.  We started with the ORCA~\cite{VectorBloxORCA} soft RISC-V processor
augmented with Lightweight Vector Extensions
(LVE)~\cite{VectorBloxLVE}.\footnote{ORCA implements a pipelined RV32IM
instruction set.  LVE is proprietary.} LVE streams data from a dedicated
scratchpad through the RISC-V
ALU, enabling efficient vector and matrix operations without any loop, memory
access, or address generation overhead.  Since LVE allows custom ALUs to be
inserted, we added the CNN accelerator, as well as a quad-16b to 32b SIMD add,
and a 32b-to-8b activation function.  These latter two custom instructions
allow us to avoid overflows but maintain performance by accumulating 16b
convolutions into 32b sums every 16 input maps, and ultimately produce an 8b
activation.

The entire system, shown in Figure~\ref{fig:system}, uses a Lattice iCE40
UltraPlus Mobile Development Platform (MDP).  The scratchpad is built from
single-ported 128kB RAM; this operates at 72MHz to provide two reads and one
write every 24MHz CPU clock.  Operating concurrently with the CPU, a DMA engine
transfers multiple 32b values from the SPI Flash ROM, which stores the binary
weights (about 270kB), into the scratchpad.  A VGA-resolution camera
($640\times 480$ pixels) using RGB565 colour is downscaled to $40\times 30$
pixels in hardware, and uses DMA to write 32b-aligned RGBA pixels into the
scratchpad.  Software de-interleaves these pixels into three separate colour
planes, padding each plane with black to $40\times 34$ pixels.  Convolutions
are computed over a $32\times 32$ centred region.

\begin{figure}[t]
    \centerline{\includegraphics[width=1.1\columnwidth]{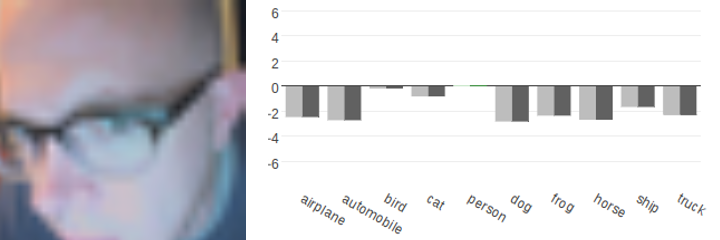}}
    \centerline{\includegraphics[width=1.1\columnwidth]{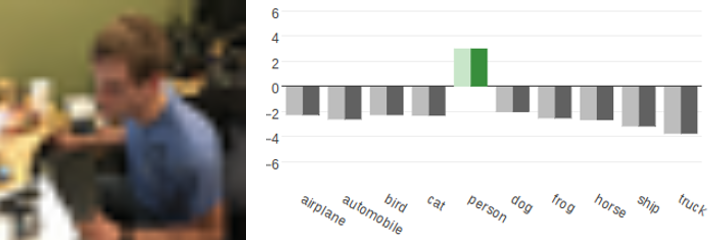}}
    \centerline{\includegraphics[width=1.1\columnwidth]{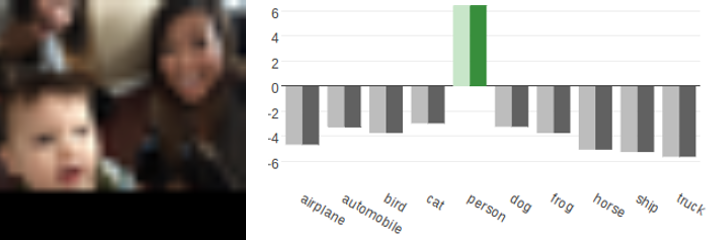}}
    \caption{Person detection, sample results \label{fig:person}}
\end{figure}

We created a person detector by training a 10-category classifier with a
modified CIFAR-10 dataset, replacing the `deer' images with duplicated images
from the `people' superclass of CIFAR-100.  In the sample results of
Figure~\ref{fig:person}, classifier scores are shown for the ten categories
using floating-point (left) and 8b fixed-point (right) activations.  A more
positive score is better.

For better performance, we reduced the network structure further and trained a
new 1-category classifier using a proprietary database of 175,000 faces and
non-face images.  

\section{Results}

On the MDP, the 10-category classifier runs in \tencatruntime.
The CPU runs at 24MHz, using 4,895 (of 5,280)
4-input LUTs, 4 (of 8) DSP blocks, 26 (of 30) 4096b BRAM, and
all four 32kB SPRAM in the Lattice iCE40 UltraPlus-5K FPGA.  The accelerator
improves ORCA RISC-V runtime of convolution layers $73\times$, and LVE improves
runtime of dense layers $8\times$, for an overall speedup of $71\times$.
In comparison, a 4.00GHz Intel i7-4790k desktop, using Python/Lasagne, takes \inteltencatruntime.

The 1-category classifier runs in \onecatruntime\ (\intelonecatruntime\ on the
i7-4790k) with \onecaterrorint8\% error and consumes 21.8~mW.  A power-optimized version, designed to
run at one frame per second, consumes just 4.6~mW.

\end{document}